\documentclass{article}
\usepackage{amsmath, amssymb}

\renewcommand{\abstractname}{Abstract.}
\renewcommand\abstract{\hfil\break\topsep=0pt\partopsep=0pt\parsep=0pt\itemsep=0pt\relax
\trivlist\item[\hskip\labelsep
{\bfseries\abstractname}]\if!\abstractname!\hskip-\labelsep\fi}

\newcommand{\email}[1]{{(e-mail: #1)}}

\def\keywordname{{\bfseries Key words:}}
\def\keywords#1{\par\addvspace\baselineskip\noindent\keywordname\enspace
\ignorespaces#1}

\def\jelclassname{{{\bfseries JEL Classification:}} }
\def\jelclass#1{\par\addvspace\medskipamount\noindent\jelclassname\
\ignorespaces#1}

\def\subclassname{{\bfseries Mathematics Subject Classification (1991):} }
\def\subclass#1{\par\addvspace\medskipamount\noindent\subclassname\
\ignorespaces#1}

\def\title#1{\hfil\break\hfil\break
\hfil\break\par\addvspace\baselineskip\noindent
\ignorespaces{\LARGE\bf#1}\hfil\break}

\def\author#1{\par\addvspace\baselineskip\noindent
\ignorespaces{\large\bf#1}}

\def\institute#1{\par\addvspace\baselineskip\noindent
\ignorespaces{\small#1}\hfil\break}

\usepackage{enumerate}
\usepackage{amsthm}
\usepackage{graphicx}

\numberwithin{equation}{section}
\newtheorem{thm}{Theorem}[section]

\newtheorem{lem}[thm]{Lemma}

\newcommand{\eq}[1]{(\ref{#1})}

\newcommand{\lemm}[1]{Lemma \ref{#1}}
\newcommand{\theor}[1]{Theorem \ref{#1}}

\newcommand{\sect}[1]{Section \ref{#1}}

\newcommand{\bbe}{\begin{equation}}

\newcommand{\R}{{\bf R}}

\newcommand{\cP}{{\mathcal P}}
\newcommand{\cC}{{\mathcal C}}

\newcommand{\eps}{\epsilon}
\newcommand{\de}{\delta}

\newcommand{\al}{\alpha}
\newcommand{\be}{\beta}

\newcommand{\ee}{\end{equation}}

\newcommand{\la}{\lambda}
\newcommand{\lp}{{\lambda_+}}
\newcommand{\lm}{{\lambda_-}}

\newcommand{\sg}{\sigma}

\newcommand{\INT}{\int_{-\infty}^{+\infty}}

\begin{document}
\title{The American put and European
options near expiry, under L\'evy processes}

\author{Sergei Levendorski\v{i}}

\institute{Department of Economics, The University of Texas at
Austin, 1 University Station C3100, Austin, TX 78712-0301, U.S.A.
 \email{leven@eco.utexas.edu}}





\begin{abstract}
We derive explicit formulas for time decay, $\theta$, for the
European call and put options at expiry, and use them to calculate
 analytical approximations to the price of the American put and early exercise boundary
 near expiry. We show that for many families of non-Gaussian
 processes
used in empirical studies of financial markets, the early exercise
boundary for the American put without dividends  is separated from
the strike price by a non-vanishing margin on the interval $[0,
T)$.
  As the riskless rate vanishes and the drift decreases
accordingly so that the stock remains a martingale,  the optimal
exercise price goes to zero uniformly over the interval $[0, T)$.
The implications for parameters' fitting are discussed.
\end{abstract}

\keywords{L\'evy processes, early exercise boundary}

\jelclass{D81, C61}

\subclass{90A09, 60H30, 60G44}


\section{Introduction}
\setcounter{equation}{0} By now, jump-diffusion processes and more
general L\'evy processes or  L\'evy driven processes are widely
used in models of the stock dynamics and term structure models,
under both historic and risk neutral measure. However, only
discrete time series of empirical data are available, therefore
the problem of identification of the jump and diffusion components
is a non-trivial task. The problem of the identification of the
diffusion component was considered in detail in A{\i}t-Sahalia
(2002, 2003); in the present paper, the focus is on the jump
component. Clearly, the presence of jumps is most prominent in
prices of contingent claims near expiry, especially for out-of-the
money options, and therefore it is natural to extract information
about the jump part of the process by using empirical data for
out-of-the-money European options. Carr and Wu (2003) showed that
in the presence of jumps, the prices of out-of-the-money European
options near expiry are of order $\tau$, where $\tau$ is the time
to expiry, and use this observation to test for the presence of
jumps. However, the results in Carr and Wu (2003) are qualitative
rather than quantitative.

The principal aim of the present paper is to derive simple
analytical formulas which can be used for the identification of
the jump part of a L\'evy process (under a risk-neutral measure
chosen by the market); the straightforward generalizations to
mean-reverting processes driven by L\'evy processes (the
Ornstein-Uhlenbeck type processes), stochastic volatility models
with jumps and L\'evy-driven term structure models will be
published elsewhere.  The first group of results of the present
paper are formulas for $\theta$, or time decay, of out-of-the
money European call and put options on a stock, at expiry.
Contrary to the Gaussian case, $\theta$ for the former (resp., the
latter)  is shown to be negative if there are positive (resp.,
negative) jumps. In other words,
 we prove that near expiry, these prices  are linear functions of
time to expiry (up to relatively small errors), and calculate the
coefficients. The results are explicit analytical expressions in
terms of the parameters which characterize the process. By using
these formulas and empirical data for options near expiry as in
Carr and Wu (2003), it is possible to identify the jump part of a
particular L\'evy model. After that, by using the data for options
far from expiry, one can identify the first instantaneous moment
of the process, and calculate the diffusion coefficient by
subtracting the second instantaneous moment of the jump part of
the process. This gives a new identification method for the
diffusion component.

By using the formulas for the $\theta$'s for the European put and
call, we obtain approximations for the American put price and
early exercise boundary near expiry. The pricing of American
options in the Gaussian case is well understood -- see e.g.
Musiela and Rutkowski (1997), Karatzas and Shreve (1998), Carr and
Faguet (1994), Carr (1998), Longstaff and Schwartz (2001),
Cl\'ement et al. (2002), and the bibliography therein. In the
infinite horizon case, explicit analytical formulas are obtained,
and in the finite horizon case, a number of efficient numerical
methods are developed. Although explicit formulas are not
available in the finite horizon case, several general results are
proved for this case as well. Consider the American put with the
strike price $K$ and maturity $T$, on a non-dividend paying stock;
the riskless rate $r>0$ is constant. Let $H(t)=H(r, K, T; t)$ be
the optimal exercise price of the American put. If the stock
log-price $X_t=\log S_t$ follows the Brownian motion, then the
asymptotic behavior of $H(r, K, T; t)$ is known both near expiry:
\begin{equation}\label{asbgauss}
\log(H(r, K, T; t)/K)\sim -\sg\sqrt{(t-T)\ln(T-t)},
\end{equation}
as $t\to T-0$ (see, e.g., Barles {\it et al.} (1995) and Lamberton
(1995); an asymptotic formula for the put price is also
available),
 and for small values of $r$: for any
$t<T$,
\begin{equation}\label{asymprgauss}
H(r, K, T; t)\to 0,\quad {\rm as}\ r\to+0
\end{equation}
(see, e.g. Musiela and Rutkowski (1997);  \eq{asymprgauss} is
formulated as follows: if the riskless interest rate is zero, then
the early exercise of the American put is not optimal).
 Notice that \eq{asbgauss} implies that
 \begin{equation}\label{exlim}
 \lim_{t\to T}H(r, K, T; t)=K,
 \end{equation}
and that \eq{exlim} and \eq{asymprgauss} do not agree well when
both the riskless interest rate and time to expiry vanish.

 In Levendorski\v{i} (2003), a variant of
Carr's randomization method  was developed for wide families of
L\'evy processes, and as a by-product, it was shown that for many
families of non-Gaussian processes used in empirical studies of
financial markets, the analogs of \eq{exlim} and \eq{asymprgauss}
agree much better. Namely,
\begin{equation}\label{exlimlevy}
 \lim_{t\to T}H(r, K, T; t)=H_T(r, K),
 \end{equation}
 where $H_T(r, K)$ depends on some of parameters which
 characterize the price process, and is {\em smaller} than the strike
 price, $K$. Moreover, it was proved that $H_T(r, K)$  vanishes with $r$, and therefore,
\begin{equation}\label{asymprlevy}
H(r, K, T; t)\to 0,\quad {\rm as}\ r\to 0, \hskip0.2cm uniformly\
in \hskip0.2cm t\in [0, T).
\end{equation}
Notice that \eq{asymprlevy} looks more natural than
\eq{asymprgauss}, and may be of some importance in today's almost
deflationary world. Further, if the difference $K-H_T(r, K)$ is
sizable, then any numerical procedure or parameters' fitting
procedure which does not take this fact into account, may produce
significant errors near expiry and strike. Hence, the non-standard
behavior of the early exercise boundary has practical implications
for the parameters' fitting and development of accurate numerical
methods.

However, the proof in Levendorski\v{i} (2003) has three
disadvantages, which stem from the use of the Wiener-Hopf method.
First, the non-standard behavior of the early exercise boundary
results from long calculations without clear intuition. Next, the
approximate formulas for the put price are expressed in terms of
the factors in the Wiener-Hopf factorization formula, hence, they
are complicated for practical use, and finally, the method in
Levendorski\v{i} (2003) is not directly applicable to a popular
class of Variance Gamma Processes, which was used in Carr et al
(2002), Madan and Hirsa (2003), and Carr and Hirsa (2003) for
pricing of American put. The generalization of the method for more
general classes of processes seems impossible.

 Here we prove \eq{exlimlevy} and \eq{asymprlevy} by using
a much simpler argument. We obtain an efficient approximation for
the put price near expiry, and formulate the result about the
non-standard behavior of the early exercise boundary in a more
meaningful form. Also, we extend the result in Levendorski\v{i}
(2003) to Variance Gamma Processes, and in addition,
 we  find a condition on a process which ensures that
the behavior of the boundary is even more regular than in the
Gaussian case: there exists a negative constant $h^0$ s.t.
\begin{equation}\label{superregular}
H(r, K, T; t)-K\sim h^0\cdot (T-t)<0,\quad {\rm as}\ t\to T.
\end{equation}
We prove that the limit of the boundary at expiry is located where
$-\theta$ of the European call is equal to the riskless rate. (If
the $-\theta$ of the European call at expiry is less than the
riskless rate for all spot prices less than the strike, then the
limit is the strike.) When the corresponding equation is solved,
one can use the formulas for $\theta$'s of the European call and
put and obtain an efficient approximation to the put price near
expiry. The formulas are fairly simple and can be used for
parameters' fitting purposes near expiry, both for the European
options and American put. They also can be used to improve
performance of numerical methods: when using the backward
induction, one can start not at expiry, where the value function
is most irregular but close to it, and use the analytical
approximation as the basis for the backward induction.

We demonstrate how large the deviations from the standard behavior
of the early exercise boundary and put price can be by using  the
empirical data for the American put documented in Carr et al.
(2002). We found that the non-standard behavior should be observed
for reasonable values of the riskless rate: typically, the margin
between the strike and boundary is 10\% or more, and for the spot
price in the 5\%--interval around the strike,   the ``jump
premium" in the put price over the pay-off is $0.5\tau$ or more,
where $\tau$ is the time to expiry (with the normalization
$\tau=1$ for a business year).

The last aim of the paper is to describe the restrictions on the
log-price process under an EMM, which should be satisfied if the
non-standard behavior of the early exercise boundary is observed.
As \eq{exlim}, \eq{exlimlevy} and \eq{superregular} imply, various
types of the behavior of $H(r, K, T; t)$ near expiry are possible.
However,  for many families of processes used in empirical studies
of financial markets, namely, Hyperbolic Processes, Normal Inverse
Gaussian processes (NIG), and Truncated L\'evy Processes of the
extended Koponen's family (a.k.a. KoBoL processes, a.k.a. CGMY
model) of order $\nu\in (1, 2)$, the behavior of the early
exercise boundary is non-standard for all parameters' values. This
means that if in the empirical studies, the early exercise
boundary for put options on a particular stock exhibits the
standard behavior near expiry, then the use of any of the
processes above to model the log-price dynamics under a
risk-neutral measure chosen by the market is inconsistent with the
data. Normal Inverse Gaussian processes, Hyperbolic Processes and
KoBoL processes of order greater than 1 are processes with sample
paths of infinite variation. Notice that the non-standard behavior
of the early exercise boundary is proved for much wider classes of
infinite variation L\'evy processes as well. For processes with
finite variation jump component, the prominent examples being
Variance Gamma Processes and KoBoL processes of order less than 1,
both the non-standard behavior \eq{exlimlevy} and ``super-regular"
behavior \eq{superregular} of the early exercise boundary are
possible, depending on parameters of the process; there are also
borderline cases, when we are able only to prove that \eq{exlim}
holds. According to a recent empirical study in Carr et al.
(2002), finite variation processes fit empirical data better.

Even for a finite variation process, if we fix all the parameters
of a process except for the drift $\mu$, let $r\to 0$ and change
$\mu$ accordingly so that the stock remains a martingale, then
below a certain critical value of $r$ (depending on the parameters
of the process), the non-standard behavior is observed, and as $r$
decreases further, the optimal exercise price goes to zero
 uniformly over the interval $[0, T)$.

It is natural to ask how robust the result about the non-standard
behavior of the early exercise boundary is. In Levendorski\v{i}
(2003) and the present paper, it is assumed that dynamics of
log-prices $X_t=\log S_t$ follows a non-Gaussian L\'evy process
under a risk-neutral measure chosen by the market. This simplest
choice of a non-Gaussian risk-neutral measure for pricing of the
American put was employed in Carr et al (2002), Madan and Hirsa
(2003), Carr and Hirsa (2003), where parameters' fitting and
computational issues where addressed. Better fit can be achieved
with stochastic volatility models with jumps but the method in
Levendorski\v{i} (2003) cannot be generalized for stochastic
volatility models.  The method of the present paper can be applied
to stochastic volatility models (and term structure models), and
under certain parameter restrictions, the margin between the early
exercise boundary and strike can be derived. Further, in
Levendorski\v{i} (2003) and the present paper, the formulas for
the margin are derived in two different approximate models, and in
cases which were considered in the both papers, the formulas for
the margin are the same. The identical results in two
approximations allows one to hope that the margin in the
continuous time model is the same.  As a proven theorem, we can
only claim that the margin in the approximate model is not less
than in the continuous time model.

The rest of the paper is organized as follows. In \sect{levy}, we
recall the definition of L\'evy processes, and give examples of
several families of  L\'evy processes used in empirical studies of
financial markets. In \sect{europe},  $\theta$'s of
out-of-the-money European put and call options are calculated, and
in  \sect{amerput}, the approximate formulas for the American put
price and early exercise boundary near expiry are derived. In
\sect{implications}, we consider several families of L\'evy
processes used (and suggested for use) in empirical studies of
financial markets, and discuss the implications of non-standard
behavior of prices and early exercise boundary for the choice of a
family of processes, and parameters' fitting in more detail. In
\sect{empirics}, we derive properties of the early exercise
boundary for KoBoL processes, for parameters' values documented in
Carr et al. (2002), and show that the margin between the strike
price and early exercise boundary can be significant indeed: more
than 10\%.  In the appendix, the technical proofs are given.

\section{L\'evy processes}\label{levy}
Recall that a L\'evy process is a
process with stationary independent increments (for general
definitions, see e.g. Sato (1999)). A L\'evy process may have a
Gaussian component and/or pure jump component. The latter is
characterized by the density of jumps, which is called the L\'evy
density. We denote it by $F(dx)$. Also, a L\'evy process can be
completely specified by its characteristic exponent, $\psi$,
definable from the equality $E\left[e^{i\xi
X(t)}\right]=e^{-t\psi(\xi)}$ (we confine ourselves to the
one-dimensional case). If $X_t$ is a L\'evy process with finite
variation jump component, then the characteristic exponent is
given by
\begin{equation}\label{psi1}
\psi(\xi)=-i\mu \xi+\frac{\sg^2}{2} \xi^2+\INT(1-e^{i\xi y})F(dy),
\end{equation}
where $\sg^2$ and $\mu$ are the variance and drift coefficient of
the Gaussian component, and $F(dy)$ satisfies
\[ \int_{\R\setminus \{0\}} \min\{1, |y|\}F(dy)<+\infty. \]
 Equation \eq{psi1} is a special
case of the L\'evy-Khintchine formula; for the general case, see
e.g. Sato (1999).

Wide families of jump-diffusion processes used in the theoretical
and empirical studies of financial markets are L\'evy processes
with finite variation jump component.

\vskip0.1cm\noindent {\it Example 2.1.} Let $X$ be a L\'evy
process with the L\'evy density
\begin{equation}\label{densjump}
F(dx)=c_+\lp e^{\lp x}{\bf 1}_{(-\infty, 0)}(x) dx + c_-(-\lm)
e^{\lm x}{\bf 1}_{(0, +\infty)}(x) dx,
\end{equation}
 where $\lp>0>\lm$, and
$c_\pm> 0$. Then
\begin{equation}\label{jumpdif}
\psi(\xi)= \frac{\sg^2}{2}\xi^2-i\mu\xi+\frac{ic_+\xi}{\lp+i\xi}
+\frac{ic_-\xi}{\lm+i\xi},
\end{equation}
where $\sg^2> 0$ and $\mu\in \R$ are the variance and drift of the
Gaussian component. The $\psi(\xi)$ is holomorphic in the strip
$\Im\xi\in (\lm, \lp)$. In order that $E[e^{X_t}]$ be finite, we
need to impose an additional condition $\lm<-1$. For the sake of
brevity,  we will consider below the exponential jump-diffusions
with the characteristic exponent of the form \eq{jumpdif}. The
generalization to the case of L\'evy densities given by general
exponential polynomials on positive and negative half-axis is
straightforward.

\vskip0.1cm\noindent {\it Example 2.2.} Variance Gamma Processes
(VGP) have been developed and used by Madan and co-authors in a
series of papers during 90th (see Madan et al. (1998)
and the bibliography there). The characteristic exponent of a VGP
without the diffusion component can be represented in the form
\begin{equation}\label{chexpvgp}
\psi(\xi)=-i\mu\xi+c[\ln(\lp+i\xi)-\ln\lp+\ln(-\lm-i\xi)-\ln(-\lm)],
\end{equation}
where $\lp>0>\lm, c>0$ and $\mu\in\R$. The condition
$E[e^{X_t}]<+\infty$ imposes an additional restriction $\lm<-1$.

\vskip0.1cm\noindent {\it Example 2.3.} Truncated L\'evy Processes
(TLP) constructed by Koponen (1995) were used for modeling in real
financial markets in Bouchaud and Potters (1997), Cont et al.
(1997) and Matacz (2000); a generalization of this family was
constructed in
Boyarchenko and Levendorski\v{i} (1999, 2000) and called the
extended Koponen family of TLP processes. Later, this
generalization was used in Carr et al. (2002) under the name
CGMY-model. As A.N. Shiryaev remarked, the name TLP was
misleading, and so starting with Boyarchenko and Levendorski\v{i}
(2002a-2002c), we use the name KoBoL processes.

The characteristic exponent of a KoBoL process  is of the form
\begin{equation}\label{kbl1d}
\psi(\xi)=-i\mu\xi+c\Gamma(-\nu)[\lp^\nu-(\lp+i\xi)^\nu+(-\lm)^\nu-(-\lm-i\xi)^\nu],
\end{equation}
where $\nu\in (0, 2), \nu\neq 1, c>0, \lm<0<\lp$, and $\mu\in \R$.
The condition $E[e^{X_t}]<+\infty$ imposes an additional
restriction $\lm\le -1$. (In Boyarchenko and Levendorski\v{i}
(1999, 2000, 2002b), the reader can also find a formula for the
KoBoL process of order $\nu=1$).

 For $\nu\in
(0,1)$, the equation \eq{kbl1d} is obtained from \eq{psi1} with
the L\'evy density given by
\begin{equation}\label{denskbl} F(dx)=c\lp e^{\lp x}{\bf 1}_{(-\infty, 0)}(x)|x|^{-\nu-1} dx
+ c(-\lm) e^{\lm x}{\bf 1}_{(0, +\infty)}(x)x^{-\nu-1} dx,
\end{equation} and  $\sg=0$ (that is, there is no
Gaussian component). In the case  $\nu\in(1, 2)$, instead of
\eq{psi1}, the general form of the L\'evy-Khintchine formula is
needed (see Sato (1999) and Boyarchenko and Levendorski\v{i}
(2002b)). KoBoL processes of order $\nu\in (0, 1)$  are
finite-variation processes, and the ones of order $\nu\in (1,2)$
are infinite-variation processes.

Other examples of infinite variation processes are Hyperbolic
Processes (HP), Normal Inverse Gaussian processes (NIG) and Normal
Tempered Stable (NTS) L\'evy processes of order $\nu \in (1,2)$;
in the general classification of Boyarchenko and Levendorski\v{i}
(1999, 2000, 2002a-2002c), HP and NIG are processes of order 1.
Hyperbolic Processes were constructed and used by Eberlein and
co-authors (see e.g. Eberlein and Keller (1995), Eberlein et al.
(1998), Eberlein and Raible (1999)); hyperbolic distributions were
constructed in Barndorff-Nielsen (1977). Normal Inverse Gaussian
processes (NIG) were introduced in Barndorff-Nielsen (1998) and
used to model German stocks in Barndorff-Nielsen and Jiang (1998);
generalization of the class NIG, namely, the class of Normal
Tempered Stable (NTS) L\'evy Processes, was constructed in
Barndorff-Nielsen and Levendorski\v{i} (2001) and
Barndorff-Nielsen and Shephard (2001). As KoBoL processes, NTSLP
can be of any order between 0 and 2.

\vskip0.1cm \noindent
 {\it Example 2.4.} The
characteristic exponent of an NTS L\'evy process is of the form
\begin{equation}\label{nig1d}
\psi(\xi)=-i\mu\xi+\de[(\al^2-(\be+i\xi)^2)^{\nu/2}-(\al^2-\be^2)^{\nu/2}],
\end{equation}
where $\de>0,$   $\al>|\be|$, and $\nu \in (0, 2)$. The condition
$E[e^{X_t}]<+\infty$ imposes an additional restriction
$-\al+\be\le-1$. With $\nu=1$, we obtain the characteristic
exponent of NIG processes.

\vskip0.1cm\noindent {\it Remark 2.1.} For the explicit
calculations in the next two sections, it is important that the
characteristic exponent of any of L\'evy processes considered
above admit the analytic continuation into the complex plane with
cuts $(-i\infty, i\lm], [i\lp, +i\infty)$, where $\lm\le -1$ and
$\lp>0$, and the analytic continuation is defined by the same
formula (in Example 2.4, $\lm=-\al+\be$ and $\lp=\al-\be$, and in
Example 2.1, the characteristic exponent is analytic in the
complex plane with two poles at $i\lm$ and $i\lp$). This property
is unnecessary if we want to obtain only qualitative results. For
instance, the existence of a non-vanishing margin between the
strike and early exercise boundary will be proved for more general
class of L\'evy processes with the jump part of infinite
variation.

\section{Time decay of out-of-the-money European call and put
options, at expiry}\label{europe} To simplify the presentation, we
normalize the strike price to 1. Let $X_t$ be a L\'evy process of
any class considered above.  Let  $\cC(x, \tau)$ and $\cP(x,
\tau)$ be the price of the European call and put option,
respectively, at time $t=T-\tau$, and the spot price $S_t=e^x$.
Denote by $\cC(x), x<0$, and $\cP(x), x>0$, the opposite to the
$\theta=\theta(x, 0)$ of the out-of-the-money European call and
put options, respectively, at expiry (the option's $\theta$ is the
derivative of the option price with respect to time): for $x<0$,
\begin{equation}\label{cC}
\cC(x):=\lim_{\tau\to +0}\frac{\cC(x, \tau)}{\tau},
\end{equation}
and for $x>0$,
\begin{equation}\label{cP}
\cP(x):=\lim_{\tau\to +0}\frac{\cP(x, \tau)}{\tau}.
\end{equation}
 If the process is
Gaussian, then the prices of out-of-the-money European call,
$\cC(x, \tau), x<0$, and put, $\cP(x, \tau), x>0$, vanish faster
than any power of $\tau$, as $\tau\to +0$, hence  the limits
\eq{cC} and \eq{cP} equal 0. The next theorem and lemma show that
in the case of non-Gaussian L\'evy processes which we consider,
both $\cC(x), x<0,$ and $\cP(x), x>0$, exist and do not vanish.
The intuition is that when the jumps are present, the value of
waiting of a positive (respectively, negative) movement in the
price is larger than in the Gaussian case.
\begin{thm}\label{margin1} Let $X_t$ be any of
L\'evy processes considered above. Then \vskip0.cm\noindent
 a) for $x<0$, the limit
\eq{cC} exists, and it is positive;
 \vskip0.cm\noindent b) for $x>0$, the limit
\eq{cP} exists, and it is positive;
 \vskip0.cm\noindent c) the time decay at
expiry, of out-of-the money European call, $\cC(x),$ $ x<0,$ and
put, $\cP(x), x>0$, depends only on the positive and negative jump
parts of the process, respectively, but not on the drift and
Gaussian component.
\end{thm}
\begin{proof}
Notice that the intuition behind the independence of the time
decay at expiry of the Gaussian component is simple: the movements
caused by the latter are of order $\tau^{1/2}$, on average, and
therefore the main contribution to the price of an
out-of-the-money option, (almost) at expiry, can come from the
jumps in the corresponding direction only (for a rigorous proof,
see the appendix). We start with the simplest case of a pure jump
process  (Example 2.1 with $\sg=0$). Let $L$ be the infinitesimal
generator of $X$. Then the price $f(X_t, t)$ of a contingent claim
with the terminal pay-off $g(X_T)$ can be represented in the form
\begin{equation}\label{europclaim}
f(x, t)=\exp[-\tau(r-L)]g(x+\mu \tau). \end{equation} For the pure
jump process, $L$ acts as follows:
\[ Lf(x)=\INT (f(x+y)-f(x))F(dy),
\]
and since $\INT F(dy)<+\infty$ for processes in Example 1.1, we
conclude that $L$ is a bounded operator in $L_\infty(\R)$. Hence,
if $g$ is bounded (the case of the put option: $g(x)=(1-e^x)^+$),
we can write \eq{europclaim} in the form
\begin{equation}\label{europclaim2}
f(x, t)=\sum_{j=0}^{\infty}\frac{\tau^j}{j!}(L-r)^jg(x+\mu\tau).
\end{equation}
If the put option is out-of-the-money, that is, $x>0$, then for
small $\tau$, $g(x+\mu\tau)=0$, hence the zero-order term in
\eq{europclaim2} is zero, and
\begin{eqnarray*}
f(x, t)&=&\tau (L-r)g(x+\mu\tau)+o(\tau)\\
&=&\tau Lg(x) +o(\tau).
\end{eqnarray*}
By dividing by $\tau>0$ and passing to the limit, we obtain
\[
\cP(x)=Lg(x).
\]
It remains to calculate $Lg(x)$ at $x>0$. We have
\begin{eqnarray*}
Lg(x)&=&\INT ((1-e^{x+y})^+-(1-e^x)^+)F(dy)\\
&=& \INT (1-e^{x+y})^+ F(dy)\\
&=&\int_{-\infty}^0 (1-e^y)c_+\lp e^{-\lp x+\lp y}dy,
\end{eqnarray*}
and by calculating the integral, we obtain
\begin{equation}
\label{cPjump} \cP(x)=\frac{c_+e^{-x\lp}}{1+\lp},\quad x>0.
\end{equation}
The case of the call option requires essentially the same
calculations but $L$ should be regarded as a bounded operator in a
space of continuous functions, which grow as $e^x$ as
$x\to+\infty$ (for details, see Boyarchenko and Levendorski\v{i}
(2002b)), and the result is
\begin{equation}\label{cCjump}
\cC(x)=\frac{c_-e^{-x\lm}}{-1-\lm},\quad x<0.
\end{equation}
Similarly, $\cC(x)$ and $\cP(x)$ for more general exponential
jump-diffusions can be calculated. Moreover, the proof above gives
\begin{equation}\label{cCgen}
 \cC(x)=\INT (e^{x+y}-1)^+F(dy),\quad x<0,
 \end{equation}
 and
 \begin{equation}\label{cPgen}
 \cP(x)=\INT (1-e^{x+y})^+F(dy), \quad x>0,
 \end{equation}
 provided $\int_{-\infty}^1 F(dy)<+\infty$ and
 \begin{equation}\label{denspos1} \int_1^{+\infty}e^y
F(dy)<+\infty, \end{equation} (the latter condition is necessary
for the stock to be priced).  We conjecture that \eq{cCgen} and
\eq{cPgen} are valid for any L\'evy process satisfying
\eq{denspos1}. However, we were unable to prove \eq{cC}--\eq{cP}
in the full generality, and so in the proof of \theor{margin1} for
the other families of L\'evy processes listed above, we calculate
the limits $\cC(x)$ and $\cP(x)$ directly, by using the formulas
for the characteristic exponents. In the case of NTS L\'evy
processes and NIG, set $\lm=-\al+\be, \lp=\al+\be$. Now, for any
$X_t$ of KoBoL, NTS, and VGP classes, and for $z<\lm$ and $z>\lp$,
define
\[
\Psi(z)=i[\psi(iz+0)-\psi(iz-0)].
\]
\begin{lem}\label{gen} Let $X_t$ be a VGP, KoBoL, or NTS L\'evy
process.

Then for $x<0$,
 \begin{equation}\label{genc} \cC(x)=(2\pi)^{-1}\int_{-\infty}^\lm \frac{e^{-xz}\Psi(z)}{z(1+z)}dz,
 \end{equation}
 and for $x>0$,
\begin{equation}\label{genp} \cP(x)=(2\pi)^{-1}\int^{+\infty}_\lp
\frac{-e^{-xz}\Psi(z)}{z(1+z)}dz.
 \end{equation}
\end{lem}
\begin{proof} In the appendix. \end{proof}
 In the Gaussian case, $\Psi(z)=0, \ \forall\ z$, and since
$\Psi$ depends linearly on $\psi$, we conclude that $\Psi$ depends
on the jump component of the process only. Hence, $\cC(x), x<0$,
and $\cP(x), x>0$, depend on the jump component only. The fact
that the former depends on the positive jumps only, and the latter
on the negative ones, follows by inspection of the
L\'evy-Khintchine formula: for $z>\lp$ (resp., $z<\lm$), $\Psi(z)$
is independent of the density of negative (resp., positive) jumps.


 In the appendix, we will calculate $\Psi(z)$ for each family of processes,
 and derive from \eq{genc}--\eq{genp} the following formulas:

1) for KoBoL processes of order $\nu\in (0, 2), \nu\neq 1$:
\begin{eqnarray}\label{cCkbl}
\cC(x)&=&-\frac{c\Gamma(-\nu)\sin \pi\nu}{\pi}\int_{-\infty}^{\lm}
\frac{e^{-xz}(-z+\lm)^\nu}{z(z+1)}dz,\quad x<0;\\\label{cPkbl}
\cP(x)&=&-\frac{c\Gamma(-\nu)\sin \pi\nu}{\pi}\int^{+\infty}_{\lp}
\frac{e^{-xz}(z-\lp)^\nu}{z(z+1)}dz,\quad x>0;
\end{eqnarray}
the integrals converge since $\lm\le -1<0<\lp$ and $\nu>0$;

2) for VGP:
\begin{eqnarray}\label{cCvgp}
\cC(x)&=&c\int_{-\infty}^{\lm}\frac{e^{-xz}}{z(1+z)}dz,\quad
x<0;\\\label{cPvgp}
 \cP(x)&=&c\int^{+\infty}_{\lp}
\frac{e^{-xz}}{z(1+z)}dz,\quad x>0;
\end{eqnarray}
the integrals converge since $\lm< -1<0<\lp$;

3) for NTS L\'evy processes
\begin{eqnarray}\label{cCnts}
\cC(x)&=&\frac{\de}{\pi}\sin\frac{\pi\nu}{2}\int_{-\infty}^{-\al+\be}
\frac{e^{-xz}[(z-\be)^2-\al^2]^{\nu/2}}{z(z+1)}dz,\quad x<0;\\
\label{cPnts}
\cP(x)&=&\frac{\de}{\pi}\sin\frac{\pi\nu}{2}\int^{+\infty}_{\al+\be}
\frac{e^{-xz}[(z-\be)^2-\al^2]^{\nu/2}}{z(z+1)}dz,\quad x>0;
\end{eqnarray}
with $\nu=1$, formulas for NIG processes obtain. The integrals
converge since $-\al+\be\le -1$, $\al+\be>0$, and $\nu>0$.
\end{proof}
 By direct inspection of formulas  \eq{cCkbl}--\eq{cPnts}, we
conclude that with $x=0$, the integrands decay as $|z|^{\nu-2}$ as
$z\to \pm\infty$ for processes of order $\nu\in (0, 2)$; if $X_t$
is a VGP or exponential jump-diffusion, then the integrands decay
faster than $|z|^{\eps-2}$, for any $\eps>0$. Therefore, with
$x=0$, the integrals diverge if and only if the jump component of
$X_t$ is a process of order $\nu\in [1, 2)$. Further, the
integrands are positive monotone continuous functions of $x$. By
using the Monotone Convergence Theorem, we deduce
\begin{thm}\label{margin2}
a) For processes of order $\nu\in (0, 1)$, VGP, and exponential
jump-diffusion processes in Example 2.1, there exist finite limits
\begin{eqnarray}\label{limc}
\cC(-0):&=&\lim_{x\to -0}\cC(x),\\\label{limp}
\cP(+0):&=&\lim_{x\to +0}\cP(x).
\end{eqnarray}

b) For  NIG, and KoBoL and NTS L\'evy processes of order $\nu\in
(1, 2)$, the limits $\cC(-0)$ and $\cP(+0)$ exist but are
infinite.

c) $\cC$ is an increasing continuous function, which maps
$(-\infty, 0)$ onto $(0, \cC(-0))$, and $\cP$ is a decreasing
continuous function, which maps  $(0, +\infty)$ onto $(\cP(+0),
0)$.
\end{thm}
{\em Remark 3.1.} a) It is possible to derive formulas for
$\cC(x), x<0,$ and $\cP(x), x>0$, for the case of Hyperbolic
Processes, and that part b) of \theor{margin2} holds in this case.

b) Part b) can be proved for more general classes of L\'evy
processes. For instance, $\cC(-0)=+\infty$ can be proved if in a
neighborhood of 0, the density of positive jumps, $F^+(dy)$,
admits a lower bound via $cy^{-2}dy$, where $c>0$. Then  there
exists a density $F^+_2(dy)$ such that $\int_0^{+\infty}e^y
F^+_2(dy)<+\infty$ and
\[
F^+(dy)\ge  F^+_1(dy) + F^+_2(dy), \] where
$F^+_1(dy)=cy^{-2}e^{-y}$. Clearly,  $\cC(x)$ cannot increase if
we replace $F^+(dy)$ with $F^+_1(dy) + F^+_2(dy)$. The argument in
the proof of \theor{margin1} for the jump-diffusion case shows
that the contribution of $F^+_2(dy)$  to $\cC(x)$ is bounded
uniformly in $x<0$, and the proof for KoBoL processes shows that
$F^+_1(dy)$ gives a contribution which is unbounded as $x\to-0$.

\vskip0.1cm

To finish this section, we list approximate formulas for European
call and put options near expiry, which follow from
\theor{margin1} and the put-call pairity. Of course, these
approximations are too inaccurate in a very small neighborhood of
the strike if the process is of order $\nu \ge 1$.

For $x>0$:
\begin{eqnarray}\label{putp}
\cP(x, \tau)&\sim & \tau \cP(x);\\\label{callp}
 \cC(x, \tau)&\sim&
e^x-1+\tau(r-\cP(x));
\end{eqnarray}
for $x<0$:
\begin{eqnarray}\label{callm}
\cC(x, \tau)&\sim & \tau \cC(x);\\\label{putm}
 \cP(x, \tau)&\sim&
1-e^x+\tau(r-\cC(x)).
\end{eqnarray}

\section{Early exercise boundary for the American put}\label{amerput}
\subsection{Finite variation processes}\label{finvar}
Assume that $\cC(-0)$ and $\cP(+0)$ exist. They are measures of
the influence of positive and negative
 jumps, respectively, and the EMM condition is naturally formulated in terms of
 $\sg, \mu, r$ and $\cC(-0)$ and $\cP(+0)$. As it was shown above,
 the limits (finite or infinite) exist for the families of
 processes in Examples 2.1--2.4.
 \begin{lem}\label{emm}
 Let $X_t$ be a L\'evy process with finite variation jump
 component. Then $e^{-rt}e^{X_t}$ is a local martingale if and only if
 \begin{equation}\label{emm1}
 r-\mu-\frac{\sg^2}{2}=\cC(-0)-\cP(+0).
 \end{equation}
 \end{lem}
 \begin{proof}  Since the Gaussian component $\sg B_t$ and
 jump component $Y_t$ of a L\'evy process are
 independent, $e^{-rt}e^{X_t}$ is a local martingale if and only if for any
 $s$ and $t>0$,
 \begin{eqnarray*}
e^{X_{s}}&=&e^{(\mu-r)t+X_s}E\left[e^{\sg
 B_t}\right]E\left[e^{Y_t}\right]\\
 &=&e^{(\mu-r+\sg^2/2)t+X_s}E\left[e^{Y_t}\right],
 \end{eqnarray*}
 whereupon
 \[
 e^{(r-\sg^2/2-\mu)t}=E\left[e^{Y_t}\right],
 \]
 and
\[
 e^{(r-\sg^2/2-\mu)t}-1=E\left[e^{Y_t}\right]-1,
 \]
 which is
\[
 e^{(r-\sg^2/2-\mu)t}-1=\INT p_t(y)(e^y-1)dy.
 \]
 (Here $p_t$ denotes the transition kernel of the semigroup
 $\{T_t\}_{t\ge 0}$,
 \[\left.T_tf(y):=E[f(Y_t)\ |\
 Y_0=y]=\int_{-\infty}^{+\infty}p_t(z)f(y+z)dz.\right)
 \]
 Divide by $t$:
\[
t^{-1}( e^{(r-\sg^2/2-\mu)t}-1)=\int_0^\infty t^{-1}
p_t(y)(e^y-1)dy-\int_{-\infty}^0 t^{-1}p_t(y)(1-e^y)dy,
 \]
 and pass to the limit $t\to 0$; the result is \eq{emm1}.
 \end{proof} We rewrite \eq{emm1} as
 \begin{equation}\label{emm2}
 \mu-\cP(+0)=r-\cC(-0)-\sg^2/2;
 \end{equation}
 for processes without the Gaussian component,
 \eq{emm2} is simpler:
\begin{equation}\label{emm3}
 \mu-\cP(+0)=r-\cC(-0).
 \end{equation}
The first theorem below shows that if negative jumps are present
but their influence is not very large, and  compensated by a
positive drift (which clearly decreases the value of waiting),
then the early exercise boundary converges to the strike price.
Moreover, the convergence is more regular than in the Gaussian
case in the sense that \eq{superregular} holds.

Let $h_*(\tau)$ be the early exercise boundary. The continuous
time model can be approximated by a discrete time model with small
time steps $\Delta=T/n$. In the discrete time approximation, the
option can be exercised at $t=T, T-\Delta, T-2\Delta,\ldots$ only.
Denote by $H:=H_{T-\Delta}(r, K)$ the optimal exercise boundary at
the last moment before the expiry, and set $h=\log H$. One should
expect that the behavior of $h=h(\Delta)$ for small $\Delta$ is a
good proxy for the behavior of the optimal exercise boundary near
expiry. In particular,
\begin{equation}\label{just}
\lim_{\Delta\to+0}h(\Delta)=\lim_{\tau\to+0}h_\ast(\tau),
\end{equation}
if the limit in the LHS exists. The proof of a simplified version
\begin{equation}\label{just2}
\lim_{\Delta\to+0}h(\Delta)\ge \lim_{\tau\to+0}h_\ast(\tau)
\end{equation}
is straightforward: as we make the class of admissible stopping
times smaller, the price of the American put does not increase,
therefore the early exercise boundary does not go down. We were
unable to prove the equality \eq{just} but the inequality
\eq{just2} suffices to make the conclusion that if the early
exercise boundary in the discrete time approximation exhibits the
non-standard behavior in the limit $\Delta\to+0$, then the
behavior of the early exercise boundary in the continuous time
model is also non-standard.
\begin{thm}\label{extrastan}
Let $X_t$ be a VGP or KoBoL process of finite variation,  without
Gaussian component, or an exponential jump process. Let
\begin{equation}\label{cneg}
\mu-\cP(+0)>0.
\end{equation}
Then
\begin{equation}\label{formneg}
\lim_{\Delta\to 0}\frac{h(\Delta)}{\Delta}=-\cP(+0).
\end{equation}
\end{thm}
In other words, {\em if $-\theta$ of the out-of-the-money European
put, at the expiry and strike, is less than the drift, then the
slope of the boundary near expiry equals the $-\theta$.}
\begin{proof}
 At $S_{T-\Delta}=H$, a put owner must be indifferent between
exercising and holding the put, therefore $h=h(\Delta)$ is the
solution to the equation
 \[
 1- e^h= e^{-r\Delta}\INT p_\Delta (y)(1-e^{h+\mu\Delta+y})^+dy
 \]
 (the LHS is the value of the put if it is exercised, and the RHS
 is the expected present value of the put if it is kept alive), equivalently,
\begin{equation}\label{exb1}
 1- e^h= e^{-r\Delta}\int_{-\infty}^0 p_\Delta
 (y-h-\mu\Delta)(1-e^{y})dy.
 \end{equation}
Assume that for small $\Delta>0$, $h(\Delta)+\mu\Delta>0$, then
$0\ge h(\Delta)>-\mu\Delta$, hence $h(\Delta)\to 0$. By using the
Taylor formula and dividing by $\Delta$, we obtain
 \begin{equation}\label{eqneg2}
 -\frac{h}{\Delta}=\int_{-\infty}^0 \Delta^{-1}p_\Delta
 (y-h-\mu\Delta)(1-e^{y})dy+O(\Delta).
\end{equation} Then we
pass to the limit $\Delta\to 0$
\[
\lim_{\Delta\to 0}\frac{-h(\Delta)}{\Delta}=\cP(+0).
\]
Condition $\cP(+0)\le\mu $ is necessary lest the assumption
$h(\Delta)+\mu\Delta\ge 0$ lead to a contradiction; and the
condition \eq{cneg} suffices for the inequality
$h(\Delta)+\mu\Delta\ge 0$ to hold for small $\Delta$.
 \end{proof}
 Now assume that \eq{cneg} fails. This happens when either the drift is negative or
 the influence of negative jumps is too large. Hence,
 the value of waiting should increase, and the optimal exercise
 boundary should be lower. As we will see, apart from the
 borderline case $\mu=\cP(+0)$,  the value of waiting becomes so large that
 $h=h(\Delta)$ remains bounded away from 0, as $\Delta\to 0$.
\begin{thm}\label{thm:finvar}
Let $X_t$ be a VGP or KoBoL process of finite variation,  without
Gaussian component, or an exponential jump process.

Then a) if $\mu=\cP(+0)$, then the limit
\begin{equation}\label{barh}
\bar h=\lim_{\Delta\to 0}h(\Delta)
\end{equation}
exists, and it is equal to 0;

b) if
\begin{equation}\label{cneg2}
\mu-\cP(+0)<0,
\end{equation}
then the limit \eq{barh} exists. It is negative, and can be found
from the equation
\begin{equation}\label{formpos1}
r=\cC(\bar h),
\end{equation}
which has a unique solution.
\end{thm}
In other words, {\em if $-\theta$ of the out-of-the-money European
put, at the expiry and strike, is bigger than the drift, then the
limit of the early exercise boundary at expiry is less than the
strike. It is located where $-\theta$ of the out-of-the-money
call, at expiry, is equal to the riskless rate.}
\begin{proof} We start with part b). First of all, we notice that due to \eq{emm3}, \eq{cneg2}
is equivalent to
\begin{equation}\label{cpos2}
\cC(-0)>r,
\end{equation}
therefore on the strength of \theor{margin2}, c), the solution
$\bar h$ to \eq{formpos1} exists and it is unique. Secondly, the
reader may be surprised that the answer is formulated in terms of
the density of positive jumps, but the proof shows that this is a
consequence of the put-call pairity (for European options). Since
$e^{-rt+X_t}$ is a (local) martingale under the risk-neutral
measure, we have
\[
\INT p_\Delta(y)e^{x+\mu\Delta+y}dy=\INT
p_\Delta(y-x-\mu\Delta)e^{y}dy=e^{x+r\Delta}.
\]
Hence, we can rewrite the RHS in \eq{exb1} as
 \begin{eqnarray*}
&& e^{-r\Delta}\left(\INT p_\Delta
 (y-\mu\Delta-h)(1-e^{y})dy
 -\int_{0}^{+\infty} p_\Delta (y-\mu\Delta-h)(1-e^{y})dy\right)\\
 &&=e^{-r\Delta}-e^h+e^{-r\Delta}\int_{0}^{+\infty} p_\Delta(y-\mu\Delta-h)(e^{y}-1)dy
 \end{eqnarray*}
 (essentially, this is the put-call pairity), and
 therefore, \eq{exb1} can be written as
\begin{equation}\label{exb2}
e^{r\Delta}-1=\int_{0}^{+\infty}
p_\Delta(y-\mu\Delta-h)(e^{y}-1)dy.
 \end{equation}
 We see that if $\mu\Delta+h(\Delta)$ is negative for small $\Delta$,
 then $h$ depends only on the upper tail of the
 probability density. By dividing by $\Delta$ and passing to the
 limit, we obtain \eq{formpos1}. Thus, part b) has been proved.

 In part a), $\mu=\cP(+0)$, hence from  \eq{emm1}, $r=\cC(-0)$, and if we assume that for some
 sequence $\Delta_n\to 0$, the sequence $h(\Delta_n)$ remains
 bounded away from 0, then by dividing by $\Delta=\Delta_n$ in \eq{exb2}
 and passing to the limit, we obtain an inequality, contradiction.
 \end{proof}
In the following theorem, we allow for a Gaussian component.
\begin{thm}\label{gauss}
 Let $X_t$ be a VGP or KoBoL of order $\nu\in (0, 1)$, or
exponential jump-diffusion process, with a Gaussian component.

Then a) if $r>\cC(-0)$, then the limit $\bar h$ exists. It is
negative, and can be found from \eq{formpos1};

b) if $r\le \cC(-0)$, then   the limit $\bar h$ exists, and it is
equal to 0.
\end{thm}
\begin{proof}
By repeating the proof of
 \theor{thm:finvar}, we obtain a), and similar arguments
 show that under condition $r\le \cC(-0)$, an assumption that the behavior of the boundary
 is non-standard, leads to contradiction. This gives b).
 \end{proof}
We see that in the case $r\le \cC(-0)$, the result is weaker than
in the pure jump case, where we can single out the case when the
boundary is more regular than in the Gaussian case.

\subsection{Infinite variation processes}\label{infvar}
According to part b) of \theor{margin2}, now $\cC(x)\to +\infty$
as $x\to -0$. By using this observation, and repeating the proof
of part a) of \theor{thm:finvar}, we obtain
\begin{thm}\label{thm:infvar}
Let $X_t$ be a NIG or KoBoL or NTSL process of infinite variation.

Then the limit $\bar h=\lim_{\Delta\to 0}h(\Delta)$ exists. It is
negative, and can be found from the equation \eq{formpos1}.
\end{thm}
{\em Remark 4.1.} If the density of positive jumps satisfies the
conditions in Remark 3.1 b), then $\cC(-0)=+\infty$, and repeating
the proof of \theor{thm:finvar}, we obtain that $\bar h$ is
negative.

\subsection{Optimal exercise price when the riskless rate
vanishes}\label{vanishing} Let $X_t$ be a process from any of
families in Examples 2.1--2.4. Let $r\to +0$ but the L\'evy
density and the diffusion coefficient of the Gaussian component
remain fixed. Then the drift of the process changes with $r$ and
converges to a finite value. By the direct inspection of
\eq{formpos1}, we conclude that if the early exercise boundary
exhibits the non-Gaussian behavior \eq{exlimlevy} near expiry,
then the early exercise price near expiry vanishes with $r$ as
well. The same argument as in the Gaussian case shows that for any
$\tau>0$, $h_*(\tau)=h_*(\tau, r)\to +0$ as $r\to +0$ (see
\eq{asymprgauss}), therefore we conclude that the optimal exercise
price tends to zero uniformly on the interval $[0, T)$.

Notice that if $X_t$ is a VGP or a KoBoL process of order $\nu\in
(0, 1)$, or exponential jump-diffusion (as in Example 2.1),  then
for large $r$, \eq{formpos1} has no solution, but for sufficiently
small $r$, equation \eq{formpos1} has a unique solution, and this
solution tends to 0 together with $r$. This means that for
sufficiently low levels of the riskless rate, all the processes
considered above should lead to the non-standard behavior of the
early exercise boundary.

\subsection{Approximate formulas for the American put price near
expiry} By using the approximate formulas for the European put and
call near expiry at the end of Section 3, we obtain approximate
formulas for the American put price near expiry:
\begin{equation}\label{amputapprox}
v(x, \tau)\sim \left\{ \begin{array}{lll} 1-e^x, & x\le \bar h\\
1-e^x+\tau(\cC(x)-r), & \bar h\le x<0 \\
\tau\cP(x), & x>0
\end{array}\right.
\end{equation}
Notice that this approximation is too inaccurate very close to the
strike, especially for infinite variation processes.

\section{Implications for parameters' fitting}\label{implications}
The behavior of the early exercise boundary near expiry can be
used to determine the type of a L\'evy process, which can be used
to model the log-price dynamics under the risk-neutral measure.
This choice of a risk-neutral measure was  used in Carr et al.
(2002), Hirsa and Madan (2003) and Carr and Hirsa (2003).
Empirical studies of financial markets clearly indicate the
presence of both downward and upward jumps (under historic
measure), therefore if the boundary exhibits the standard
behavior, then one may not use simultaneously a process of order
$\nu\in [1,2)$ under the historic measure and L\'evy process under
a risk-neutral measure (chosen by the market). Indeed, as it is
shown in Carr et al. (2002), p.312, the difference between the
historic and risk-neutral L\'evy densities must be of finite
variation, therefore a process of infinite (finite) variation
under the historic measure must remain a process of infinite
(finite) variation under an EMM. Processes of order $\nu<1$ can be
used for modelling of log-price dynamics both under historic and
risk-neutral measure but additional restrictions on the density of
the upward jumps must be imposed. Another possibility is to use
one-sided KoBoL processes with densities of positive jumps of an
order different from the one for positive jumps (see Boyarchenko
and Levendorski\v{i} (1999, 2000, 2002b)). The characteristic
exponents of these strongly asymmetric KoBoL processes are of the
form
\begin{eqnarray*}
\psi(\xi)&=&-i\mu\xi+c_+\Gamma(-\nu_+)[\lp^{\nu_+}-(\lp+i\xi)^{\nu_+}]\\
&&\hskip1cm
+c_-\Gamma(-\nu_-)[(-\lm)^{\nu_-}-(-\lm-i\xi)^{\nu_-}],
\end{eqnarray*}
where $c_\pm\ge 0$, $\nu_\pm<2, \lm<0<\lp$, and $\nu_\pm\neq 1$
(for the formulas in the case $\nu_\pm=1$, see Boyarchenko and
Levendorski\v{i} (1999, 2000, 2002b)). The analogs of \eq{cCkbl}
and \eq{cPkbl} are
\begin{eqnarray*}\label{cCkblas}
\cC(x)&=&-\frac{c_-\Gamma(-\nu_-)\sin
\pi\nu_-}{\pi}\int_{-\infty}^{\lm}
\frac{e^{-xz}(-z+\lm)^{\nu_-}}{z(z+1)}dz,\quad
x<0;\\\label{cPkblas} \cP(x)&=&-\frac{c_+\Gamma(-\nu_+)\sin
\pi\nu_+}{\pi}\int^{+\infty}_{\lp}
\frac{e^{-xz}(z-\lp)^\nu_+}{z(z+1)}dz,\quad x>0.
\end{eqnarray*}
If a finite variation process is chosen for modelling of the
American put on a particular stock, then the choice of parameters
of the process is constrained by the observed behavior of the
early exercise boundary:
\begin{enumerate}[(i)]
\item
if the non-standard behavior \eq{exlimlevy} is observed, then the
constraint is $r<\cC(-0)$;
\item
if the ``super-regular" behavior \eq{superregular} is observed,
then the constraint is $\mu>\cP(+0)$ (and probably, there is no
Gaussian component); however, as empirical examples in the next
section show, this possibility is unlikely to realize;
\item
if $\bar h=0$ but the ``super-regular" behavior \eq{superregular}
is not documented, then the constraint is $r\ge \cC(-0)$, and the
Gaussian component is admissible.
\end{enumerate}
 Still another
possibility is to assume that under an EMM chosen by the market,
the rate of decay of the density of positive jumps is very large;
then $\bar h=0$ for processes of order $\nu<1$, and $\bar h$ is
negative but very close to 0 for processes of order $\nu\ge 1$.
For KoBoL, this means that  the steepness parameter $\lm$ must be
large in modulus. If under the historic measure, the $\lm$ is not
large, this interpretation presumes that the agents in the market
are very risk averse.

The formulas for $\theta$ of out-of-the money European call and
put options, at expiry, can also be used to infer the type of an
appropriate family of processes, and  parameters of the process.
In particular, if $\theta$ at expiry quickly grows as the spot
price approaches the strike  price, then processes with infinite
variation jump component must be used, otherwise the processes
with finite variation jump component.

\section{Empirical examples}\label{empirics}
The aim of this section is to demonstrate how large the margin
between the strike and early exercise boundary and the ``jump
premium" in the American option price over the payoff $1-e^x$ can
be for realistic values of parameters. We use risk-neutral
parameters' values of the KoBoL processes with a diffusion
component, which were obtained for several stocks in Carr et al.
(2002) (Table 3 on p.327 in op. cit.).
\begin{figure}
 \scalebox{0.6}
{\includegraphics{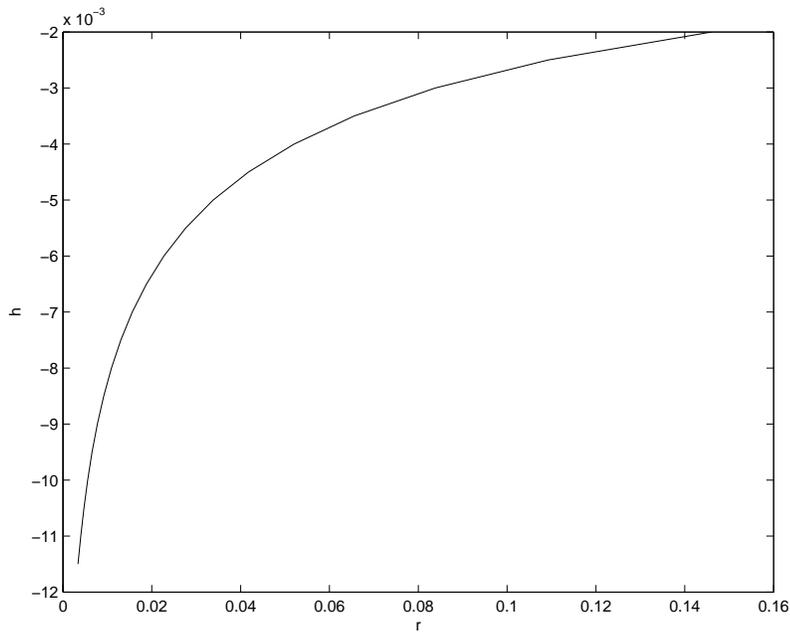}}
 \caption{Limit of log of the early exercise boundary at expiry, as a function of the riskless rate.
 Parameters (ibm1111):
  $\nu=1.0102;
c=0.42; \lp=4.37; \lm=-191.20$, $\sg=0.428$.}
 \end{figure}
 \begin{figure}
 \scalebox{0.6}
{\includegraphics{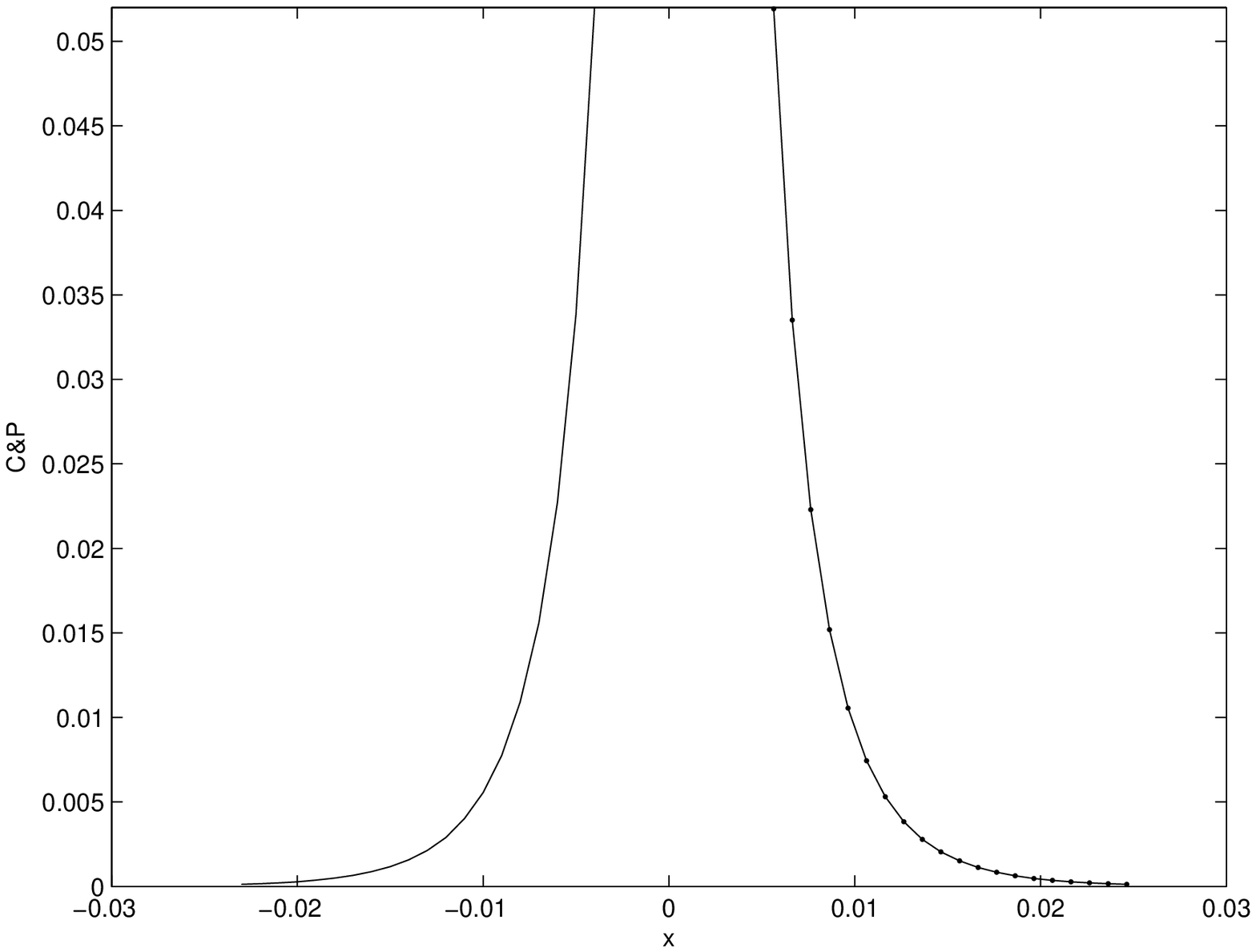}}
 \caption{Time decay of the out-of-the-money call and put,
 at expiry. Parameters (ibm1111):
  $\nu=1.0102;
c=0.42; \lp=4.37; \lm=-191.20$, $\sg=0.428$.}
 \end{figure}

\vskip0.1cm {\it Example 5.1.} Only in one case: ibm1111, the
process is of order $\nu>1$: the risk-neutral parameters are
 $\nu=1.0102;
c=0.42; \lp=4.37; \lm=-191.20$, and $\sg=0.428$ (in Carr et al.,
these parameters are denoted $Y, C, G, -M$, and $\eta$).
 Since
$\nu>1$, the non-standard behavior of the early exercise boundary
is guaranteed but the gap between the strike price and boundary is
small unless the interest rate is very small indeed. It is seen
from Figure 1, where we plot the graph of $\bar h$ as a function
of $r$, that for $r=0.02$, $\bar h$
 is about $0.006$, and for $r=0.005$, $\bar h$ is close to 0.01,
 so the gap is clearly seen.
 In Figure 2, we
 plot the graph of $\theta$ at expiry, for out-of-the-money  European
 call and put (graph of $\cC(x)$, for small $x<0$ near zero, and put, $\cP(x)$, for $x>0$).
  Both are unbounded
 in a neighborhood of $0$.
 Since the density of positive jumps decays very fast (the
 steepness parameter $\lm=-191.20$ is very large), the time decay of
  out-of-the-money puts is very small unless the
 spot price is very close to the strike. The density of positive jumps decays
 faster than the one of negative jumps ($-\lm>\lp$), therefore the $\theta$
 for the
 out-of-the-money call is smaller than the one for
 out-of-the-money put, for the same distance from the strike.

\vskip0.1cm
 {\em Example 5.2.} For amzn1014, the parameters are $\sg=0.0684;
\nu=0.3072; c=4.60; \lp=1.78; \lm=-6.29.$ The fairly large $c$
indicates that the intensity of jumps is large, and relatively
small values of $\lp$ and $\lm$ imply that there are many large
jumps, especially negative ones (recall that $\lp$ characterizes
the rate of decay of the density of negative jumps). Therefore,
even not very close to expiry,  the owner of an out-of-the-money
European option can expect that with a non-negligible probability
the option will become in-the-money. Hence, the price of these
out-of-the-money options is sizable, especially the one of the
put. Figure 3 demonstrates this effect. We plot  graphs of $\cC$
and $\cP$ in a neighborhood of zero.
\begin{figure}
 \scalebox{0.6}
{\includegraphics{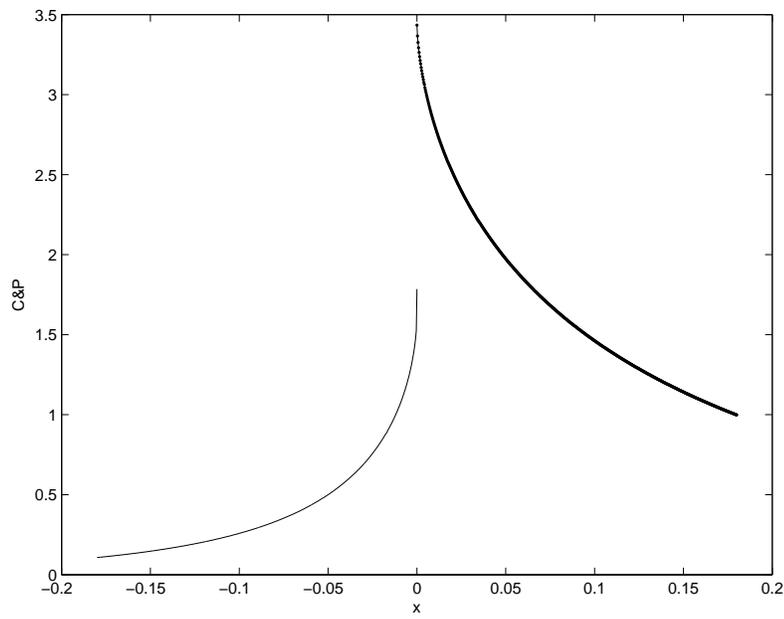}}
 \caption{Time decay of out-of-the-money European call and put,
 at expiry. Parameters (amzn1014):
  $\nu=0.3072; c=4.60; \lp=1.78; \lm=-6.29, \sg=0.0684.$}
 \end{figure}
Further, $\cC(-0)=1.7851$, therefore for reasonable $r$, we have
$r<\cC(-0)$, and the behavior of the early exercise boundary is
non-standard. Since the intensity of large negative jumps is
significant, the value of waiting is very large, and the gap
between the strike and optimal exercise boundary is also large.
From \eq{amputapprox} and Figure 3, we can conclude that even at
5\% below the strike, the option value of waiting (the difference
between the American put price and the pay-off $1-e^x$) is of
order $0.5\tau$; $\tau=1$ is one year.

In the next two Figures 4 and 5, we plot the graph of $\cC$
farther from the strike, and the graph of $\bar h$ as the function
of $r$.
\begin{figure}
 \scalebox{0.6}
{\includegraphics{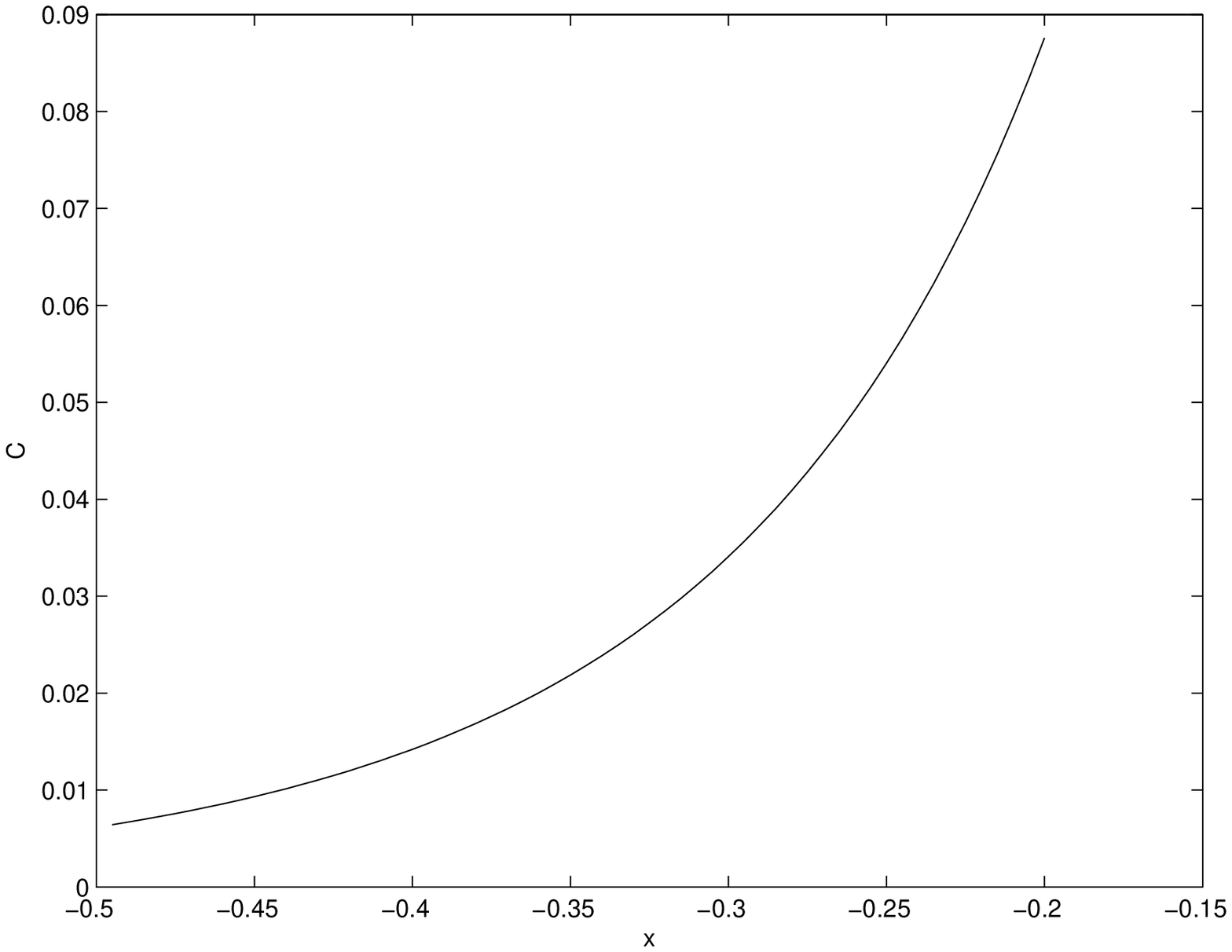}}
 \caption{Time decay of  out-of-the-money European call,
 at expiry. Parameters (amzn1014):
  $\nu=0.3072; c=4.60; \lp=1.78; \lm=-6.29, \sg=0.0684.$}
 \end{figure}
\begin{figure}
 \scalebox{0.6}
{\includegraphics{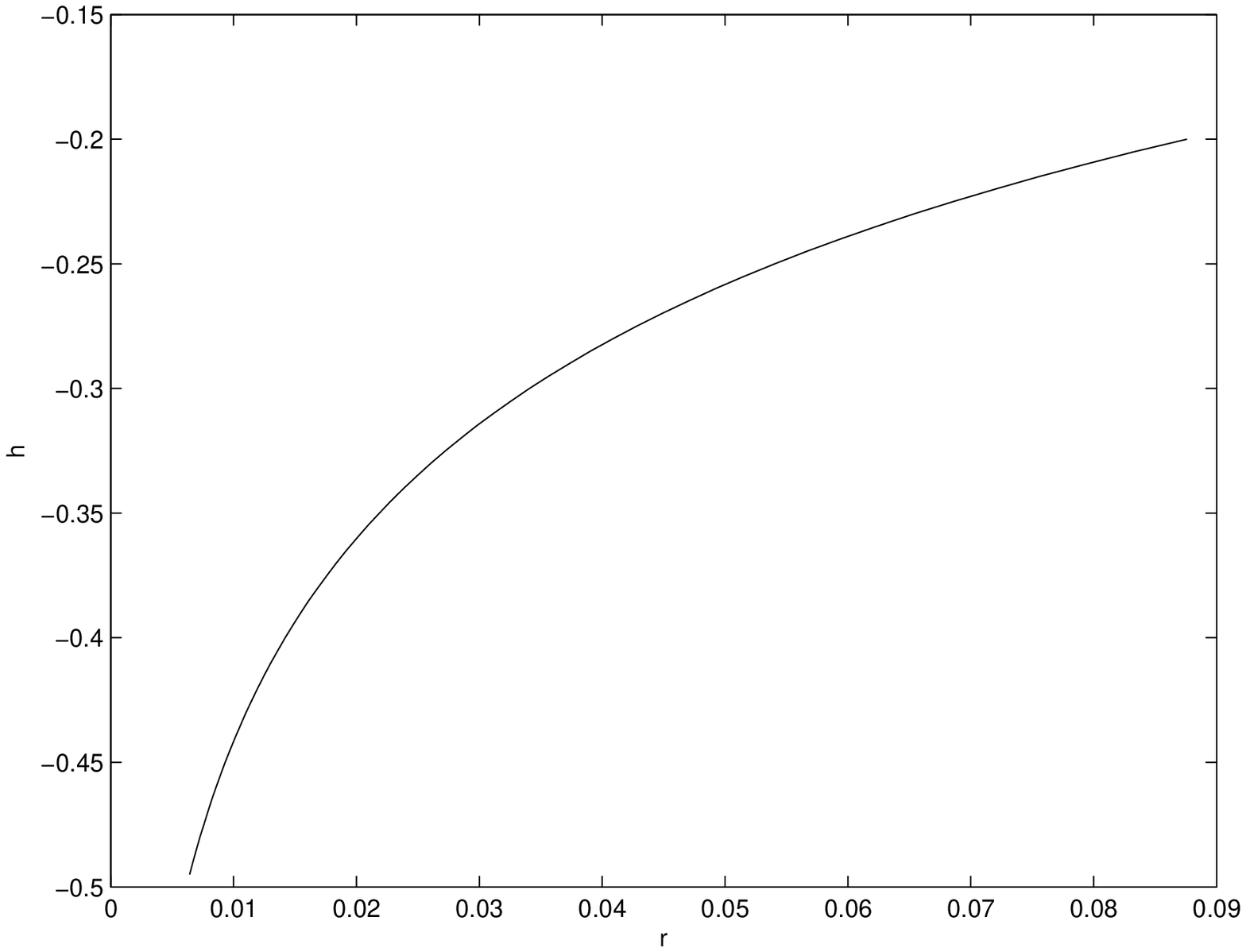}}
 \caption{Limit of log of the early exercise boundary, at expiry. Parameters (amzn1014):
  $\nu=0.3072; c=4.60; \lp=1.78; \lm=-6.29, \sg=0.0684.$}
 \end{figure}
In Figure 6, we plot both the early exercise boundary, at expiry,
$H=\exp(\bar h(r))$, as a function of $r$, and that of the early
exercise boundary far from expiry, $H_\ast=\exp(h_\ast(r))$ (the
early exercise boundary in the infinite time horizon case).
\begin{figure}
 \scalebox{0.6}
{\includegraphics{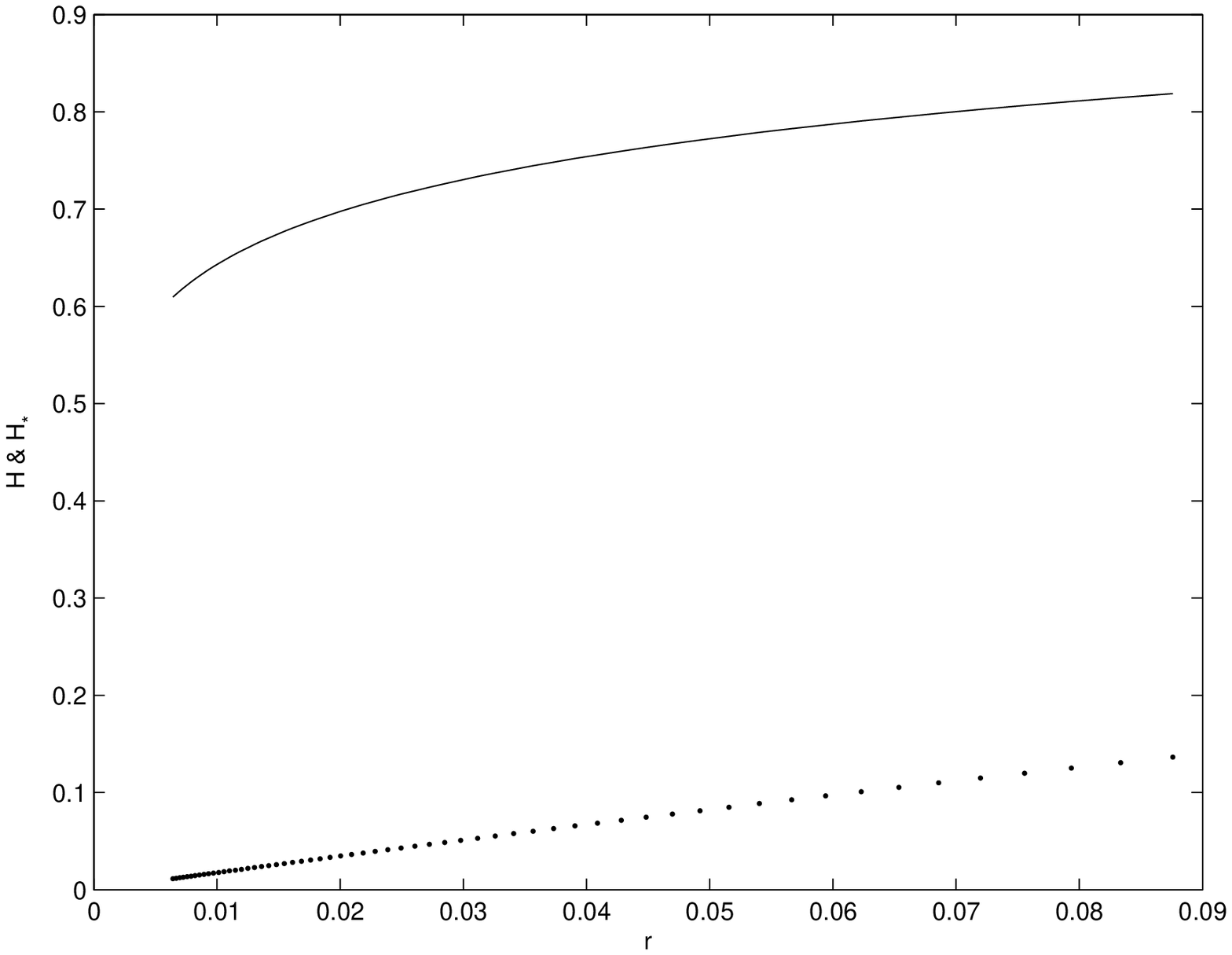}}
 \caption{Graphs of the early exercise boundary, at expiry (solid line),
 and far from expiry (dotted line), as functions of the riskless rate. Parameters (amzn1014):
  $\nu=0.3072; c=4.60; \lp=1.78; \lm=-6.29, \sg=0.0684.$}
 \end{figure}
 Similar results are valid for the other stocks documented in Carr
 et al. (2002).

\section{Conclusion} For several families of L\'evy processes used in empirical
studies of financial markets, we have proved that the time decay
of out-of-the-money European options at expiry is not negligible,
and derived explicit formulas. These formulas can be used to make
numerical calculations near expiry more accurate -- both for
European options, and American ones. By using the same formulas,
we proved that the optimal exercise boundary for the American put
is separated from the strike price by a non-vanishing margin on
the interval $[0, T)$, and that as the riskless rate vanishes, the
optimal exercise price goes to zero uniformly over the interval
$[0, T)$. The last result may be especially interesting since the
current levels of interest rates are rather low. Furthermore, we
discussed the restrictions on the process under the risk-neutral
measure which the non-Gaussian behavior of the early exercise
price near expiry imposes, and suggested a natural asymmetric
extension of the KoBoL family (a.k.a. CGMY model), which can be
used for any type of the behavior of the boundary. We showed that
for the risk-neutral parameters obtained for several stocks in
Carr et al. (2002), the behavior should be non-standard, and the
margin between the strike and the early exercise boundary be quite
sizable: more than 10\%, up to expiry. The jump premium over the
payoff is shown to be sizable as well.



\appendix
\section{Technical proofs}
\subsection{Proof of \lemm{gen}} First, we consider processes
without the Gaussian component, and in the end prove that the
results remain the same after an addition of a Gaussian component.

  Further, we assume
that $\lm<-1$; the results for the case $\lm=-1$ obtain by passing
to the limit $\lm\uparrow -1$ in the formulas for the case
$\lm<-1$. For any $\sg\in (\lm, -1)$, the price of the European
call option is given by
\begin{equation}\label{callprice}
\cC(x,
\tau)=(2\pi)^{-1}\int_{-\infty+i\sg}^{+\infty+i\sg}\frac{\exp[ix\xi-\tau(r+\psi(\xi))]}
{-i\xi(1-i\xi)}d\xi
\end{equation}
(see e.g. Boyarchenko and Levendorski\v{i} (1999, 2002b)).
Clearly, the limit $\cC(x)$ is independent of $r$, therefore we
may set $r=0$.   The characteristic exponent of a process without
the Gaussian component is of the form
\[
\psi(\xi)=-i\mu\xi+\phi(\xi), \] where $\phi(\xi)$ grows as
$|\xi|^\nu$ in the complex plane with the cuts $(-i\infty, i\lm]$
and $[i\lp, +i\infty)$, for processes of order $\nu\in (0, 2)$. in
the complex plane outside two poles $\xi=i\la_\pm$. For VGP,
$\phi(\xi)$ grows as $\ln|z|$, and  for exponential jump
processes, $\phi(\xi)$ decays as $|\xi|^{-1}$.

Set $x'=x+\mu \tau$, and write \eq{callprice} as
\begin{equation}\label{callprice2}
\tau^{-1}\cC(x,
\tau)=(2\pi)^{-1}\int_{-\infty+i\sg}^{+\infty+i\sg}\tau^{-1}\frac{\exp[ix'\xi-\tau
\phi(\xi)]} {-i\xi(1-i\xi)}d\xi.
\end{equation}
Let $X_t$ be a KoBoL or NTS L\'evy process of order $\nu\in (0,
1)$. Then we can find $\de>0$ s.t. $\rho:=(1-\de)/\nu>1$. Since
$\Re\phi(\xi)$ is bounded away from zero on the line $\Im\xi=\sg$,
the integrand in \eq{callprice2} admits an upper bound via $
C\tau^{-1}(1+|\xi|)^{-2}. $ It follows that the integral over
$|\Im\xi|\ge \tau^{-\rho}$ makes a contribution of order
\[
2 C\tau^{-1}\int_{\tau^{-\rho}}^{+\infty} \eta^{-2}d\eta=2C
\tau^{\rho-1}=o(1),\quad {\rm as}\ \tau\to+0.
\]
Since $|\phi(\xi)|\le C(1+|\xi|)^\nu$, we have
$\tau\phi(\xi)=O(\tau^\de)=o(1)$ on the set $\{\xi\ |\ \Im\xi=\sg,
|\Re\xi|\le \tau^{-\rho}\}$. Hence, we may write \eq{callprice2}
as
\begin{eqnarray}\nonumber
\tau^{-1}\cC(x,
\tau)&=&(2\pi\tau)^{-1}\int_{-\infty+i\sg}^{+\infty+i\sg}\frac{e^{ix'\xi}(1-\tau
\phi(\xi))} {-i\xi(1-i\xi)}d\xi+o(1)\\\label{au1}
 &=&(2\pi\tau)^{-1}\int_{-\infty+i\sg}^{+\infty+i\sg}\frac{e^{ix'\xi}} {-i\xi(1-i\xi)}d\xi\\\label{au2}
&&+(2\pi)^{-1}\int_{-\infty+i\sg}^{+\infty+i\sg}\frac{e^{ix'\xi}(-
\phi(\xi))} {-i\xi(1-i\xi)}d\xi+o(1).
\end{eqnarray}
The $x<0$ is fixed, therefore for sufficiently small $\tau>0$,
$x'$ is negative and bounded away from zero, and therefore,
$\Re(ix'\xi)$ is negative and bounded away from zero, uniformly in
$\tau\to+0$ and $\xi$ in the half-plane $\Im\xi\le\sg$. This means
that we can use the Cauchy theorem and push the line of the
integration in \eq{au1} down: $\sg\to-\infty$, and obtain that the
integral in \eq{au1} is zero. Hence,
\begin{equation}\label{callprice3}
\tau^{-1}\cC(x,
\tau)=(2\pi)^{-1}\int_{-\infty+i\sg}^{+\infty+i\sg}\frac{e^{ix'\xi}(-
\phi(\xi))} {-i\xi(1-i\xi)}d\xi+o(1).
\end{equation}
 By using
essentially the same argument as in Boyarchenko and
Levendorski\v{i} (2002b), where a similar integral (5.33)  was
transformed to the integral over the banks of the cut $(-i\infty,
i\lm]$ in the complex plane, and after the change of variables
$\xi=iz-0$ and $\xi=iz+0$ on the left and right banks,
respectively, to an integral over $(-\infty, \lm)$ (equation
(5.37) in Boyarchenko and Levendorski\v{i} (2002b)), we obtain,
for $x'<0$,
\begin{equation}\label{genc2}
\tau^{-1}\cC(x, \tau)=(2\pi)^{-1}\int_{-\infty}^\lm
\frac{e^{-x'z}\Psi(z)}{z(1+z)}dz+o(1).
 \end{equation}
 As $\tau\to 0$, $x'\to x$, therefore by passing to the limit in
 \eq{genc2}, we obtain \eq{genc}.

Recall that the calculations above were made for a KoBoL or NTS
L\'evy process of order $\nu\in (0,1)$. If $X_t$ is a VGP, then we
make essentially the same argument by using any $\rho>0$. If $X_t$
is a Hyperbolic Process, NIG, or KoBoL or NTS L\'evy process of
order $\nu\in (1,2)$, then we start with the integration by part
in \eq{callprice2}, by using
\[
e^{ix'\xi}d\xi=(ix')^{-1}de^{ix'\xi}.
\]
After that we obtain an absolutely converging  integral, and can
repeat the constructions above with straightforward changes. In
the end, we obtain
\[\tau^{-1}\cC(x,\tau)=
 (2\pi x')^{-1}\int_{-\infty}^\lm e^{-x'z}\left[\frac{\Psi'(z)}{z(1+z)}+
 \Psi(z)\left(\frac{1}{z(1+z)}\right)'\right]dz+o(1).
 \]
 By integrating by part back, and passing to the limit
 $\tau\to+0$, we obtain \eq{genc}.

 To finish the proof of \eq{genc}, it remains to show that
 the result does not change when we add a Gaussian component.
Let $X_t$ be any of the processes above, without a Gaussian
component, and let $X_{\sg, t}$
 be a L\'evy process with the same ``drift" term and jump density as $X_t$,
 and with the diffusion coefficient $\sg^2$. Let $p_{\sg, t}$ be
 the probability density of $X_{\sg, t}$, and $p_{G, \sg, t}$ be the
 probability density of the Brownian motion  with the volatility
 $\sg^2$.
 The $p_{G, \sg, t}$ is the convolution of $p_t$ and
 $p_{G, \sg, t}$, therefore for $x>0$,
 \begin{eqnarray*}
\tau^{-1}\cC(x, \tau)&=&\int^{+\infty}_0 \tau^{-1}p_{\sg,
 \tau}(y-x)(e^y-1)dy\\
 &=&\int^{+\infty}_0 \left(\INT \tau^{-1}p_{G, \sg,
 \tau}(z)p_\tau(y-x-z)dz\right)(e^y-1) dy\\
 &=&\INT  p_{G, \sg,
 \tau}(z) \int^{+\infty}_0\tau^{-1}p_\tau(y-x-z)(e^y-1) dy dz.
\end{eqnarray*}
Take $\rho\in (1/2, 1)$. In the region $|z|\ge \tau^{\rho}$, the
Gaussian density is $O(\tau^N)$, for any $N$, therefore we may
replace the outer integral with the integral over $z<\tau^{\rho}$,
and add $o(1)$. But then on the support of the integrand, for a
fixed $x<0$, $y-x-z$ is bounded away from zero, uniformly in
$\tau$. Define $\cC_{jump}(x)$ as $\cC(x)$ but assuming that the
log-price follows the jump part of the process. Then we can
represent the inner integral in the form
\[
\cC_{jump}(x+z)+o(1),
\]
and therefore,
\[
\tau^{-1}\cC(x, \tau)=\int_{-\infty}^{\tau^\rho} p_{G, \sg,
 \tau}(z)\cC_{jump}(x+z)dz +o(1).
 \]
 Now, define $\cC_{jump}(y)$ to be $\cC_{jump}(x/2)$ for all $y>x/2$. Then for small
 $\tau$, the integral above does not change but now $\cC_{jump}$ is
 a continuous  bounded function on $\R$. By using the super-exponential decay
 of the Gaussian density, we can restore the $\INT$, and after
 that pass to the limit and obtain $\cC(x)=\cC_{jump}(x)$.
 This finishes the proof of \eq{genc}.

 The proof of \eq{genp} is similar, only we start with the formula
 for the European put:
 \[
 \cP(x,\tau)=(2\pi)^{-1}\int_{-\infty+i\sg}^{+\infty+i\sg}
\frac{\exp[ix\xi-\tau(r+\psi(\xi))]} {-i\xi(1-i\xi)}d\xi,
\]
where $\sg\in (0,\lp)$, and push the contour of the integration
up.

\subsection{Proof of  \eq{cCkbl}--\eq{cPnts}, and \eq{cPjump}--\eq{cCjump}
for jump processes with non-zero drift} If $X_t$ is a VGP, KoBoL
or NTS L\'evy process, then it suffices to calculate $\Psi(z)$ and
substitute the results in \eq{genc} and \eq{genp}.

1. Let $X_t$ be a VGP. We have
\[
\Psi(z)=ci[\ln(\lp-z+i0)+\ln(-\lm+z-i0)-\ln(\lp-z-i0)-\ln(-\lm+z+i0)].
\]
If $z>\lp$, then the terms with $\lm$ cancel out, and
\[
\ln(\lp-z\pm i0)=\ln(z-\lp)\pm i\pi.
\]
Hence, $\Psi(z)=-c 2\pi$. If $z<\lm$, then the terms with $\lp$
cancel out, and
\[
\ln(-\lm+z\pm i0)=\ln(z+\lm)\pm i\pi.
\]
Thus, $\Psi(z)=c2\pi$.

2. Let $X_t$ be a KoBoL process. In the case $z>\lp$,
\begin{eqnarray*}
\Psi(z)&=&ic\Gamma(-\nu)[-(\lp-z+i0)^\nu-(-\lm+z-i0)^\nu\\
 && \hskip1.6cm + (\lp-z-i0)^\nu+(-\lm+z+i0)^\nu]\\
 &=&ic\Gamma(-\nu)[(\lp-z-i0)^\nu-(\lp-z+i0)^\nu]\\
 &=&ic\Gamma(-\nu)(z-\lp)^\nu i(e^{-i\pi\nu}-e^{i\pi\nu})\\
&=& 2c\Gamma(-\nu)\sin(\pi\nu)(z-\lp)^\nu.
\end{eqnarray*}
Similarly, in the case $z<\lm$,
\[
\Psi(z)=-2c\Gamma(-\nu)\sin(\pi\nu)(-z+\lm)^\nu.
\]

3. Let $X_t$ be an NTS L\'evy process. If $z>\lp=\al+\be$, then
\begin{eqnarray*}
\Psi(z)&=&i\de[(\al^2-(\be+i(iz+0)^2)^{\nu/2}-(\al^2-(\be+i(iz-0)^2)^{\nu/2}]\\
&=&i\de[(\al^2-(\be-z+i0)^2)^{\nu/2}-(\al^2-(\be-z-i0)^2)^{\nu/2}]\\
&=&i\de(\al-\be+z)^{\nu/2}[(\al+\be-z+i0)^{\nu/2}-(\al+\be-z-i0)^{\nu/2}]\\
&=&i\de
(\al-\be+z)^{\nu/2}(z-\al-\be)^{\nu/2}(e^{i\pi\nu/2}-e^{-i\pi\nu/2})\\
&=&-2\sin\frac{\pi\nu}{2}((z-\be)^2-\al^2)^{\nu/2}.
\end{eqnarray*}
If $z<\lm=-\al+\be$, then similarly, we obtain the same expression
but without the minus sign.

Finally, if $X_t$ is an exponential jump-diffusion process, then
the same proof as in \lemm{gen} shows that for the call, it
suffices to consider the pure jump process, and that in this case,
\eq{callprice3} holds. Since
\[
\phi(\xi)=\frac{ic_+\xi}{\lp+i\xi}+\frac{ic_-\xi}{\lm+i\xi},
\]
the integrand in \eq{callprice3} has the only pole $\xi=i\lm$ in
the plane $\Im\xi\le \sg$. We push down the line of integration,
and on crossing the pole, apply the residue theorem. The result is
\begin{eqnarray*}
\tau^{-1}\cC(x, \tau)&=&-(2\pi
i)^{-1}\int_{-\infty+i\sg}^{+\infty+i\sg}\frac{e^{ix'\xi}ic_-\xi}{-i\xi(1-i\xi)(\xi-i\lm)}d\xi+o(1)\\
&=& \left.\frac{e^{ix'\xi}c_-}{-1+i\xi}\right|_{i\xi=-\lm}+o(1),
\end{eqnarray*}
and by passing to the limit, we obtain \eq{cCjump}. \eq{cPjump} is
proved similarly.

\vskip1.5cm

\noindent {\bf References}

  \noindent
1. A{\i}t-Sahalia, Y.: Disentangling diffusion from jumps.
Forthcoming in Journal of Financial Economics (2003)

\noindent 2. A{\i}t-Sahalia, Y.: Telling from  discrete data
whether the underlying continuous-time model is a diffusion.
Journal of Finance 57, 2075-2112 (2002)

\noindent 3. Barles, G., Burdeau, J.,  Romano, M., and Samsoen,
N.: Critical stock price near expiration. Mathematical Finance. 5,
77-95 (1995)

\noindent 4.  Barndorff-Nielsen, O.~E.: Exponentially decreasing
distributions for the logarithm of particle size. Proceedings of
the Royal Society. London. Ser. A. 353, 401--419 (1977)

\noindent 5. Barndorff-Nielsen,  O.~E.: Processes of Normal
Inverse Gaussian Type. Finance and Stochastics. 2, 41--68 (1998)

 \noindent
6.  Barndorff-Nielsen, O.~E., and Jiang, W.: An initial analysis
of some German stock price series. Working Paper Series 15, CAF
Univ. of Aarhus/Aarhus School of Business, October 1998.

 \noindent
7. Barndorff-Nielsen, O.~E., and  Levendorski\v{i}, S.: Feller
Processes of Normal Inverse Gaussian type. Quantitative Finance.
1, 318--331 (2001)

\noindent 8. Barndorff-Nielsen, O.~E., and  Shephard, N.: Normal
modified stable processes. Forthcoming in  Theory of Probability
and Mathematical Statistics

 \noindent 9.  Bouchaud, J-P., and  Potters, M.: Theory
of Financial
  Risk. Paris:
Al\'ea-Saclay, Eurolles  1997

 \noindent 10. Boyarchenko, S.I., Levendorski\v{i}, S.Z.:
Generalizations of the Black-Scholes equation for truncated L\'evy
processes. Working Paper. University of Pennsylvania, Philadelphia
1999

\noindent 11. Boyarchenko, S.I., Levendorski\v{i}, S.Z.: Option
Pricing for Truncated L\'evy Processes. International Journal of
Theoretical and Applied Finance. 3:3, 549--552 (2000)

\noindent 12. Boyarchenko, S.I.,  Levendorski\v{i},  S.Z.:
Perpetual American Options under L\'evy Processes. SIAM Journal of
Control and Optimization. 40:6, 1663--1696 (2002a)

\noindent 13. Boyarchenko, S.I., Levendorski\v{i}, S.Z.:
Non-Gaussian Merton-Black-Scholes theory. Singapore: World
Scientific 2002b

\noindent 14. Boyarchenko, S.I., Levendorski\v{i}, S.Z.: Barrier
options and touch-and-out options under regular L\'evy processes
of exponential type. Annals of Applied Probability. 12:4,
1261--1298 (2002c)


\noindent 15. Carr, P.: Randomization and the American put. Review
of Financial Studies. 11, 597-626  (1998)

 \noindent 16. Carr, P., and  Faguet, D.: Fast
accurate valuation of American options. Working paper. Cornell
University, Ithaca 1994

\noindent 17. Carr,  P.,  Geman,  H.,  Madan, D.B., and  Yor, M.:
The fine structure of asset returns: an empirical investigation.
Journal of Business. 75, 305-332 (2002)

\noindent 18. Carr,  P., and Hirsa, A.: Why be backward?. Risk,
January 2003, 26, 103--107 (2003)

\noindent 19. Carr, P., and Wu, L.: What type of process underlies
options? A simple robust test. Journal of Finance. 57, 2581--2610
(2003)

 \noindent 20. Cl\'ement,  E.,   Lamberton, D., and
Protter, P.: An analysis of a least squares regression method for
American option pricing. Finance and Stochastics. 6, 449-471
(2002)

 \noindent 21.
 Cont,  R.,  Potters, M., and  Bouchaud, J.-P.: Scaling in
stock market data: stable laws and beyond", in Dubrulle, B.,  F.
Graner, and D. Sornette (eds.). Scale Invariance and beyond
(Proceedings of the CNRS Workshop on Scale Invariance, Les
Houches, March 1997). Berlin: Springer  1997

\noindent 22. Eberlein,  E., and  Keller, U. Hyperbolic
distributions in finance. Bernoulli. 1, 281--299 (1995)

 \noindent
23. Eberlein, E.,  Keller, U., and  Prause, K.: New insights into
smile, mispricing and value at risk: The hyperbolic model. Journal
of Business. 71, 371--406 (1998)

 \noindent
 24. Eberlein, E., and  Raible, S.: Term structure models driven
by general L\'evy processes. Mathematical Finance. 9, 31--53
(1999)

\noindent 25. Hirsa, A., and Madan, D.B.: Pricing American options
under Variance Gamma. Journal of Computational Finance,
forthcoming (2003)

 \noindent
26.  Karatzas, I., and  Shreve, S.E.:  Methods of Mathematical
Finance. Springer-Verlag: Berlin Heidelberg New York (1998)

 \noindent
 27. Koponen, I.: Analytic approach to the problem of
convergence of truncated L\'evy flights towards the Gaussian
stochastic process. Physics Review E. 52, 1197--1199 1995


\noindent 28. Lamberton, D.: Critical price for an American option
near maturity. In Bolthausen, E.~, M.~Dozzi, F.~Russo (eds.)
Seminar on Stochastic Analysis, Random Fields and Applications.
Boston Basel Berlin: Birkha\"user  1995

 \noindent
29.  Levendorski\v{i}, S.~Z.: Pricing of American put under L\'evy
processes. To appear in  International Journal of Theoretical and
Applied Finance

\noindent
 30. Longstaff, F.A., and  Schwartz, E.S.: Valuing American options by simulation:
a simple least-squares approach. Review of Financial Studies. 14,
113-148 (2001)

\noindent 31. Madan,  D.~B.,  Carr, P., and  Chang, E.~C.: The
variance Gamma process and option pricing. European Finance
Review. 2, 79--105 (1998)

 \noindent
 32. Matacz, A.: Financial modeling and option theory with
the Truncated Levy Process.  International Journal of Theoretical
and Applied Finance.  3, 143-160 (2000)

 \noindent
 33. Musiela, M., and Rutkowski, M.:  Martingale methods in
financial modelling. Berlin Heidelberg New York: Springer-Verlag
1997

 \noindent
34. Sato, K.:  L\'evy processes and infinitely divisible
distributions. Cambridge: Cambridge University Press 1999

\end{document}